# Matrix Inversion Using Cholesky Decomposition

Aravindh Krishnamoorthy, Deepak Menon
ST-Ericsson India Private Limited, Bangalore
aravindh.k@stericsson.com, deepak.menon@stericsson.com

*Abstract*—**In this paper we present a method for matrix inversion based on Cholesky decomposition with reduced number of operations by avoiding computation of intermediate results; further, we use fixed point simulations to compare the numerical accuracy of the method.**

## I. INTRODUCTION

Matrix inversion techniques based on Cholesky decomposition and the related LDL decomposition are efficient techniques widely used for inversion of positive-definite/symmetric matrices across multiple fields.

Existing matrix inversion algorithms based on Cholesky decomposition use either equation solving [3] or triangular matrix operations [4] with most efficient implementation requiring $\frac{2}{3}n^3$ operations.

In this paper we propose an inversion algorithm which reduces the number of operations by 16-17% compared to the existing algorithms by avoiding computation of some known intermediate results.

In section 2 of this paper we review the Cholesky and LDL decomposition techniques, and discuss solutions to linear systems based on them. In section 3 we review the existing matrix inversion techniques, and how they may be extended to non-Hermitian matrices. In section 4 we discuss the proposed matrix inversion method.

## II. CHOLESKY DECOMPOSITION

If $A \in C^{N \times N}$ is a positive-definite Hermitian matrix, Cholesky decomposition factorises it into a lower triangular matrix and its conjugate transpose [3], [5] & [6].

$$A = LL^* \qquad \ldots (1)$$

Or equivalently, using an upper triangular matrix $R = L^*$ as

$$A = R^*R \qquad \ldots (2)$$

In a software implementation the upper triangular matrix is preferred as operations are row-wise and compatible with C programming language.

The elements of $R = r_{ij}, i \leq j \leq N$ are given as follows.

Diagonal elements:

$$r_{ii} = \sqrt{a_{ii} - \sum_{k=1}^{i-1} r_{ki}^* r_{ki}} \qquad \ldots (3)$$

Upper triangular elements, i.e. $i < j$:

$$r_{ij} = \frac{1}{r_{ii}} \left( a_{ij} - \sum_{k=1}^{i-1} r_{ki}^* r_{kj} \right) \qquad \ldots (4)$$

Note that since older values of $a_{ii}$ aren't required for computing newer elements, they may be overwritten by the value of $r_{ii}$, hence, the algorithm may be performed in-place using the same memory for matrices $A$ and $R$.

Cholesky decomposition is of order $O(n^3)$ and requires $\frac{1}{6}n^3$ operations. Matrix inversion based on Cholesky decomposition is numerically stable for well conditioned matrices.

If $Ax = y$, with $x, y \in C^{N \times 1}$ is the linear system with $N$ variables, and $A$ satisfies the requirement for Cholesky decomposition, we can rewrite the linear system as

$$R^*Rx = y \qquad \ldots (5)$$

By letting $b = Rx$, we have

$$R^*b = y \qquad \ldots (6)$$

and

$$Rx = b \qquad \ldots (7)$$

These equations are solved using backward-substitution and require $\frac{1}{2}n^2$ operations each for the solution.

### A. LDL Decomposition

If $A \in C^{N \times N}$ is a symmetric matrix, LDL decomposition factorises it into a lower triangular matrix, a diagonal matrix and conjugate transpose of the lower triangular matrix [5].

$$A = LDL^* \qquad \ldots (8)$$

Or equivalently, using an upper triangular matrix $R = L^*$ as

$$A = R^*DR \qquad \ldots (9)$$

This decomposition eliminates the need for square-root operation.

The elements of $R = r_{ij}, i < j \leq N$, and diagonal elements of the matrix $D = diag(d_i), i \leq N$ are given as follows.

Diagonal elements:

$$d_i = a_{ii} - \sum_{k=1}^{i-1} r_{ki}^* r_{ki} d_k \qquad \ldots (10)$$

Upper triangular elements, i.e. $i < j$:

$$r_{ij} = \frac{1}{d_i}\left(a_{ij} - \sum_{k=1}^{i-1} r_{ki}^* r_{kj} d_k\right) \quad \ldots (11)$$

When efficiently implemented, the complexity of the LDL decomposition is same as Cholesky decomposition.

If $Ax = y$, with $x, y \in C^{Nx1}$ is the linear system with $N$ variables, and $A$ satisfies the requirement for LDL decomposition, we can rewrite the linear system as

$$R^*DRx = y \quad \ldots (12)$$

By letting $b = Rx$, we have

$$R^*Db = y \quad \ldots (13)$$

and

$$Rx = b \quad \ldots (14)$$

These equations are solved using backward-substitution and when efficiently implemented, require $\frac{1}{2}n^2$ operations each for the solution.

## III. EXISTING TECHNIQUES

### A. Equation Solving

If $A \in C^{NxN}$, we may find $X = A^{-1}$, $X \in C^{NxN}$ and $X = (x_1, x_2, x_3, \ldots, x_N)$ by solving

$$Ax_i = e_i \quad \ldots (15)$$

Where $e_i$ is the i[th] column of the identity matrix of order $N$ [3]. This equation may be solved using either Cholesky or LDL based method as described above depending on the properties of $A$.

In either case since $A$ is Hermitian, it is sufficient to solve for upper (or lower) half of $X$ and update the other half with the complex conjugate values as $x_{ji} = x_{ij}^*$ for $i < j \le N$.

Solving for the upper half of the matrix requires two triangular matrix solutions with $\frac{1}{3}n^3$ multiply operations each. The total number of multiply operations including the decomposition is $\frac{5}{6}n^3$.

### B. Triangular Matrix Operations

If $A \in C^{NxN}$, we may find the inverse of $A = X \in C^{NxN}$, using Cholesky decomposition, we have

$$A = R^*R \quad \ldots (16)$$

This implies:

$$A^{-1} = R^{-1}R^{*-1} \quad \ldots (17)$$

$M = R^{-1}$, and $M = (m_1, m_2, m_3, \ldots, m_N)$ is computed as follows.

$$Rm_i = e_i \quad \ldots (18)$$

Where $e_i$ is the i[th] column of the identity matrix of order $N$. Once $M$ is computed, $X$ is computed as

$$X = MM^* \quad \ldots (19)$$

When efficiently implemented, equation solving requires $\frac{1}{3}n^3$ and matrix multiplication requires $\frac{1}{6}n^3$. The total number of matrix multiplication including the Cholesky decomposition is $\frac{2}{3}n^3$ [4].

### C. Non-Hermitian Matrices

Cholesky (or LDL) decomposition may be used for non-Hermitian matrices by creating an intermediate Hermitian matrix as follows:

For an arbitrary matrix $D \in C^{NxN}$, we may construct a Hermitian matrix $A$ as $A = DD^*$. Once the inverse of A is found using Cholesky (or LDL) decomposition, we may find $D^{-1}$ as $D^{-1} = D^*A^{-1}$.

This method requires a matrix transposition operation for finding $D^*$ and matrix multiplication to find $D^{-1}$.

## IV. PROPOSED METHOD

The proposed method is a modification to the *Equation Solving* method described in section III., where the first equation's solution (6, 13) is avoided.

The method as applicable to Cholesky decomposition and LDL decomposition are described below.

### A. Using Cholesky Decomposition

If $A \in C^{NxN}$ and $X \in C^{NxN}$ such that $X = A^{-1}$, we have

$$AX = I \quad \ldots (20)$$

From Cholesky decomposition:

$$R^*RX = I \quad \ldots (21)$$

By letting $RX = B$, we have

$$R^*B = I \quad \ldots (22)$$

and

$$RX = B \quad \ldots (23)$$

Let $B = R^{*-1} = L^{-1}$, for $L = R^*$; we note that the inverse of the lower triangular matrix $L$ is lower triangular, and the diagonal entries of $L^{-1}$ are the reciprocals of diagonal entries of $L$. Therefore, the matrix B is lower triangular, with diagonal elements as reciprocals of the diagonal elements of $L$.

Now, given these properties, we construct a matrix $S$ such that

$$s_{ij} = \begin{cases} \frac{1}{l_{ii}} & \text{if } i = j \\ 0 & \text{otherwise} \end{cases} \quad \ldots (24)$$

We note that the matrix $S$ is the correct solution to upper diagonal elements of the matrix $B$, i.e. $\tilde{s}_{ij} = b_{ij}\ for\ i \le j \le N$, we do not compute B and instead use backward substitution to solve for $x_{ij}$ using the equation $Rx_i = s_i$, given:

- We solve only for $x_{ij}$ such that $i < j \leq N$ (upper triangle elements.)
- $x_{ji} = x_{ij}^*$ where needed.

Equation solving requires $\frac{1}{3}n^3$ multiply operations, the total number of multiply operations for matrix inverse including Cholesky decomposition is $\frac{1}{2}n^3$.

*B. Using LDL Decomposition*

If $A \in C^{NxN}$ and $X \in C^{NxN}$ such that $X = A^{-1}$, we have

$$AX = I \qquad \ldots (25)$$

From LDL decomposition:

$$R^*DRX = I \qquad \ldots (26)$$

By letting $RX = \tilde{B}$, we have

$$R^*D\tilde{B} = I \qquad \ldots (27)$$

and

$$RX = \tilde{B} \qquad \ldots (28)$$

From (27) we note that $\tilde{B} = (R^*D)^{-1} = L^{-1}$, for $L = R^*D$. In this case we construct the matrix $\tilde{S}$ as

$$\tilde{s}_{ij} = \begin{cases} \frac{1}{d_i} & if\ i = j \\ 0 & otherwise \end{cases} \qquad \ldots (29)$$

We note that the matrix $\tilde{S}$ is the correct solution to upper diagonal elements of the matrix $\tilde{B}$, i.e. $\tilde{s}_{ij} = b_{ij}\ for\ i \leq j \leq N$, we do not compute $\tilde{B}$ and instead use backward substitution to solve for $x_{ij}$ as in Cholesky decomposition.

Equation solving requires $\frac{1}{3}n^3$ multiply operations, the total number of multiply operations for matrix inverse including LDL decomposition is $\frac{1}{2}n^3$.

*C. Numerical Accuracy*

Fixed point simulations indicate that the proposed method has a good numerically accuracy. From figure 1 we may note that the proposed method has the lowest average norm error compared to other methods described in this paper.

## V. CONCLUSION

We presented a method for matrix inversion based on Cholesky decomposition with reduced number of operations by avoiding computation of intermediate results. The number of operations for the methods is summarised in table 1. The results of fixed point simulations are in figure 1.

| METHOD | OPERATIONS |
|---|---|
| Equation Solving | $\frac{5}{6}n^3$ |
| Triangular Matrix Operations | $\frac{2}{3}n^3$ |
| Proposed Method | $\frac{1}{2}n^3$ |

Table 1. Table summarising the number of operations for the methods described in the paper.

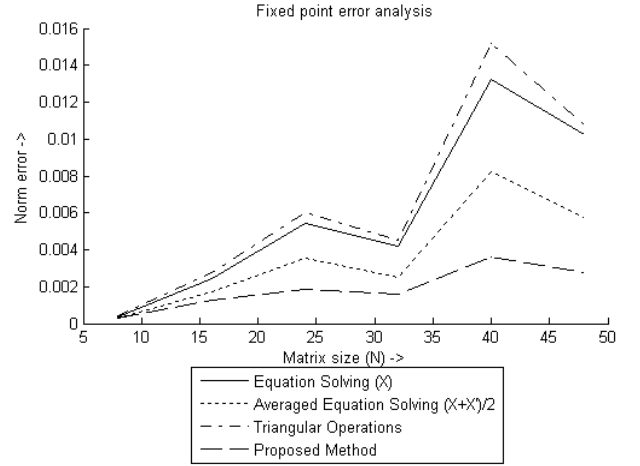

Fig. 1. Fixed point error analysis of various techniques described in the paper.